\documentclass[12pt,epsf]{article}

 \usepackage{epsfig}
 \usepackage{times}

\newcommand{\be}{\begin{equation}}
\newcommand{\ee}{\end{equation}}
\def\({\left (}
\def\){\right )}
\def\[{\left [}
\def\[{\right ]}
\def\vp{\varphi}

\begin{document}
\begin{titlepage}
\bigskip
\rightline{}
\rightline{hep-th/0505278}
\bigskip\bigskip\bigskip\bigskip
\centerline {\Large \bf {The Role of Dipole Charges in Black Hole 
Thermodynamics}}
\bigskip\bigskip
\bigskip\bigskip

\centerline{\large Keith Copsey and Gary T. Horowitz}
\bigskip\bigskip
\centerline{\em Department of Physics, UCSB, Santa Barbara, CA 93106}
\centerline{\em keith@physics.ucsb.edu, gary@physics.ucsb.edu}
\bigskip\bigskip

\begin{abstract}
Modern derivations of the first law of black holes appear to show 
that the only charges that arise are monopole charges that can be 
obtained by surface integrals at infinity. However, the recently 
discovered five dimensional black ring solutions empirically satisfy 
a first law in which dipole charges appear. We resolve this 
contradiction and derive a general form of the first law for black 
rings. Dipole charges do appear together  with a corresponding 
potential.  We also include theories with Chern-Simons terms and 
generalize the first law to  other horizon topologies and more 
generic local charges. 
\end{abstract}

\end{titlepage}

\baselineskip=16pt
\setcounter{equation}{0}

 \section{Introduction}
 
One of the first indications that there was a connection between 
black holes and thermodynamics was the discovery of the laws of black 
hole mechanics in the early 1970's \cite{Bardeen:1973gs}. Not 
surprisingly, this work was in the context of four spacetime 
dimensions.   Over the past decade there has been growing interest in 
higher dimensional black holes and black branes, so it is natural to 
ask how these laws extend. The second law extends trivially, since 
the argument that the horizon area always increases is independent 
of  spacetime dimension or horizon topology. The zeroth law also has 
an extension, since one can compute the derivative of the surface 
gravity in any dimension \cite{Racz:1995nh} (although the condition 
for it to be constant requires a field equation or symmetries). We 
will focus here on the first law, which describes how stationary 
black holes respond to small perturbations.
 
 There have been several previous derivations of the first law for 
higher dimensional black holes (see, e.g., 
\cite{Gibbons:2004ai,Korzynski:2004gr,Rogatko:2005da,Koga:2005sv}). 
However, most of these assume the horizon is topologically spherical, as in four 
dimensions. In addition, there have been several derivations which  
assume that 
the four dimensional uniqueness theorems extend to higher dimensions 
\cite{Myers:1986un,Gauntlett:1998fz,Townsend:2001rg}. It has recently 
been shown that both of these properties can be violated. There are 
five dimensional 
vacuum solutions describing stationary black rings with horizon 
topology $S^2\times S^1$ \cite{Emparan:2001wn}. These black rings 
can  have the 
same mass and angular momentum as spherical black holes. More 
importantly, in the presence of suitable matter, the nontrivial 
topology of the event horizon makes it possible for the black holes 
to carry a local dipole charge. Emparan has found a 
continuous family of  non-vacuum black  rings, all with the same 
asymptotic conserved charges and differing only by their dipole 
charge \cite{Emparan:2004wy}. 
 
 When thinking about the role of dipole charges in the first law, one 
is led to an apparent paradox. On the one hand, from the explicit 
form of the solutions, Emparan claims that the dipole charge does 
enter the first law, at least for perturbations from one stationary 
solution to another. On the other hand, a powerful and elegant 
derivation of the first law by Sudarsky and Wald 
\cite{Sudarsky:1992ty} (which does not assume black hole uniqueness) 
seems to show that the only charges that can enter into the first law 
are (monopole) charges obtained by surface integrals at infinity.

 We will review the Sudarsky-Wald argument in section 2 and 
generalize it to five dimensions. We then review the solutions found 
by Emparan in section 3  and finally resolve this paradox in section 
4. The net result is that dipole charges do appear in the general 
form of the first law in higher dimensions. In the next section, we 
extend this derivation to include a Chern-Simons term and derive a 
first law appropriate for, e.g., black rings in  minimal 5D 
supergravity. The fact that dipole charges arise in the first law  
raises the question 
of whether other charges can arise in the first law, perhaps carried 
by some not-yet-discovered black hole solution in higher dimensions. 
We discuss this in section 6, and argue that the answer is yes.  
Finally, section 7 contains some concluding remarks.

We will use Greek indices $\mu,\nu,...$ for spacetime tensors, and 
latin indices $a,b,...$ for purely spatial tensors.

\setcounter{equation}{0}
\section{Sudarsky-Wald argument for the first law}

We first consider asymptotically flat solutions of the five 
dimensional theory
\begin{equation}\label{theory}
     S = \beta \int d^5 x \sqrt{{-g}} \left [ {R} 
-\frac{1}{2}
\nabla_\mu
     \phi \nabla^\mu \phi - \frac{1}{12}e^{-\alpha
     \phi} H_{\mu\nu\rho} H^{\mu\nu\rho}\right ]
     \end{equation}
where $H = dB$ is a three form field strength, $\phi$ is 
the dilaton, $\alpha$ is the dilaton coupling, and $\beta$ is a 
normalization constant we choose to leave arbitrary for the 
present.   This is the 
simplest theory which
contains stationary black ring solutions  with dipole charge. It is 
also of interest in 
string theory and M theory. If we parameterize $\alpha$ in terms of 
an integer $N$ via $\alpha^{2} = \frac{4}{N} - \frac{4}{3}$, then for 
$N=1,2,3$ the solutions can be interpreted as arising from $N$ 
intersecting branes in higher dimensions. 
In particular, $N=1$ is the NS sector of low energy string theory (in 
the Einstein frame).
For $N=3$, the dilaton decouples and can be set to zero. In this 
case, the theory is
equivalent to Einstein-Maxwell in five dimensions by a simple duality 
transformation.

Since we have a three form in five dimensions, the natural charge 
defined at infinity is the magnetic charge 
\be\label{mcharge}
Q_M = {1\over 4\pi} \int _{S^3}  H
\ee
However, if the horizon has topology $S^2\times S^1$ one can also 
define an electric dipole charge
\begin{equation}\label{localcharge}
    q_e = \frac{1}{4\pi}\int_{S^{2}}e^{-\alpha \phi} 
\star{ H}
    \end{equation}
 where the integral is over any $S^2$ which can be continuously 
deformed to an $S^2$ on the horizon\footnote{More generally,  one 
need only require that the $S^2$ is cobordant to an $S^2$ on the 
horizon.}.  $q_e$ is well defined due 
    to the field equation $d (e^{-\alpha \phi} \star{ H}) = 
0$. 

The Sudarsky-Wald derivation of the first law is based on the  
Hamiltonian formulation of general relativity.  It was originally 
given in the context of four dimensional Einstein-Maxwell (or 
Einstein Yang-Mills) theory and goes as follows. The Hamiltonian for  
Einstein-Maxwell theory takes the
 ``pure constraint" form
\begin{equation}
H = \int_{\Sigma} (\xi^{\mu}C_{\mu} + 
\xi^{\mu}A_{\mu}\mathcal{C}) +\  {\rm surface\ terms}
\end{equation}	
where $ \Sigma $ is a spacelike surface,  $\xi^{\mu}$ is the 
time 
evolution vector field, $C_{\mu}$ are the constraints 
from the Einstein equations, and $\mathcal{C}$ is the Maxwell 
constraint   ($D_a E^a=0$).  Note we define the electric field
\begin{equation}
E^{a} = F^{\mu a}n_{\mu}
\end{equation}
with $n^\mu$ denoting the unit normal to $\Sigma$.
The surface terms are determined by the requirement that the 
variation of the Hamiltonian is well defined.  In addition to the 
usual gravitational surface terms, one gets an additional surface 
term:
\be
{1\over 4\pi} \int (\xi^{\mu}A_{\mu}) E_a dS^a
\ee
 
 Consider a stationary, axisymmetric, electrically charged black hole 
with bifurcate Killing horizon. Choose $\Sigma$ to have 
boundaries at infinity and the bifurcation surface $S$.  Let $\chi^\mu$ denote the Killing 
field which vanishes on $S$ and set $\xi^\mu = \chi^\mu$. Then the 
variation of the Hamiltonian must vanish, since this just yields the 
time derivative of the canonical variables in the direction $\chi$, 
and $\chi$ is a symmetry.\footnote{We choose a gauge with ${\cal 
L}_\chi A=0$.}  However, as long as the perturbation satisfies the 
linearized constraints, the volume term in the Hamiltonian vanishes 
by itself. This means that the sum of the variation of the surface 
terms must vanish. This yields the first law
\be
\delta M = \frac{\kappa}{8\pi} \delta A_{H} + \Omega \delta J+  
\Phi_{E} \delta Q_{E}
\ee
where $\kappa $ is the surface gravity, $A_{H}$ and $\Omega$ are, 
respectively, the area and angular velocity of the horizon
and $\Phi_{E}$ is the electrostatic potential (and we  set $G=1$). 
The origin of each term is 
the following.
Since
\be \chi = {\partial \over \partial t} + \Omega {\partial \over 
\partial \varphi}
\ee
the gravitational surface terms at infinity yield $\delta M - \Omega 
\delta J$.
 The fact that  $\xi^\mu=0$ on
$S$ implies that the only contribution from the Maxwell field comes 
from  the surface integral at infinity (assuming all fields are 
regular) and yields the $\Phi_{E} \delta Q_{E}$ term where $\Phi_{E} 
= - A_t(\infty)$. 
The gravitational surface term on $S$ does provide a nonzero 
contribution but this is only because the constraint involves the 
scalar curvature which has two derivatives of the metric. The surface 
term thus involves a derivative of $\xi^\mu$ and yields the 
$\frac{\kappa}{8\pi} \delta A_{H}$ 
term.

It is easy to generalize this to the five dimensional theory 
(\ref{theory}). The first step is to do a Hamiltonian decomposition 
of this theory.
We denote the Lie derivative of a
     tensor in the $\xi$ direction by a dot:
     \begin{equation}
	 \dot{{B}} = \mathcal{L}_{\xi} {B}
	 \end{equation}
 The momentum canonically conjugate to the spatial metric ${h}_{ab}$ 
is, as usual
\begin{equation}\label{piG}
\pi_{G}^{ab} = \frac{\partial \mathcal{L}}{\partial
\dot{{h}}_{ab}} = \beta \sqrt{{h}}
({K}^{ab} - {h}^{ab} {K} )
\end{equation}
where ${K}^{ab}$ is the extrinsic curvature and ${K}= 
{K}^{ab}{h}_{ab}$.
The momentum conjugate to the dilaton $\phi$ is
\begin{equation}\label{piphi}
\pi_{\phi} = \frac{\partial \mathcal{L}}{\partial
\dot{\phi}} = \beta \sqrt{{h}}
{n}^{\mu} \nabla _{\mu} \phi
\end{equation}
 while the momentum 
conjugate to the 2-form potential $ {B}$ is
\begin{equation}
\pi_{B}^{ab} = 
\frac{\partial
\mathcal{L}}{\partial
\dot{B}_{ab}} = \frac{\beta \sqrt{{h}}}{2} e^{-\alpha \phi} H^{ab\mu} 
n_\mu
\end{equation}
In addition to the usual gravitational constraints, there is the 
additional constraint
\be\label{Bconstraint}
D_a \({\pi_{B}^{ab}\over \sqrt h}\) =0
\ee
where  $D_{a}$ is the derivative 
compatible with spatial metric $h_{ab}$.

The general form of the Hamiltonian will be given in section 6, but 
here we simply quote the surface terms coming from the matter fields. 
In addition to the usual gravitational surface terms, we obtain
 $$ - \beta \int dS_{b} \Big [ \Big(N 
D^{b} \phi + N^{b} 
\frac{\pi_\phi}{\beta \sqrt{h}}\Big) \delta \phi - 2 \xi^\mu 
 B_{\mu c} \delta \Big(\frac{ \pi_{B}^{bc}}{\beta \sqrt{h}}\Big) 
\nonumber
$$
\begin{equation}\label{bdyterms}
+ \Big( \frac{N e^{-\alpha \phi}}{2} 
{H}^{bcd} +\frac{3}{\beta \sqrt{h}}
N^{[b}\pi_{B}^{cd]} \Big) \delta B_{cd} \Big]
\end{equation}
where $N,N^a$ are the usual lapse and shift decomposition of the 
evolution vector $\xi^\mu$.

Suppose there exists a stationary, axisymmetric solution with 
bifurcate Killing horizon.
In five dimensions, one can have rotation in two orthogonal planes.  
If there are two rotational Killing fields, the null Killing field on 
the horizon takes the general form
\be\label{defchi}
 \chi = {\partial \over \partial t} + \Omega_\varphi {\partial \over 
\partial \varphi}
 +\Omega _\psi  {\partial \over \partial \psi}
\ee
Choosing $\xi^\mu = \chi^\mu$ and 
assuming the metric is asymptotically flat in the sense that it 
approaches flat space at the same rate as the Myers-Perry black hole 
\cite{Myers:1986un}, the gravitational surface terms at infinity 
yield $\delta M-\Omega_\varphi \delta J^\varphi
-\Omega_\psi \delta J^\psi$.   On the horizon, they yield the usual 
$\frac{\kappa}{8\pi} \delta A_{H}$ term.
The main object for us is determining the possible contributions from 
the matter fields.

Since $N$ and $N^a$ both vanish on $S$, one does not expect any 
contribution from the horizon. To evaluate the contribution at 
infinity, we must be more specific about  the asymptotic behavior of 
the fields. We require the solutions to have finite energy and hence 
$T_{\mu\nu}n^{\mu}n^{\nu} = \mathcal{O}(r^{-4 - 2\epsilon} )$.   At 
leading order $T_{\mu\nu}n^{\mu}n^{\nu}$ is given by 
    a sum of positive definite terms and hence we get the following 
    restrictions:
    \begin{equation}
	H^{t r \theta_{1}} = \mathcal{O}(r^{-3 - \epsilon})
\end{equation}
     \begin{equation}\label{falloff}
	H^{t \theta_{1} \theta_{2}} = \mathcal{O}(r^{-4 - \epsilon})
\end{equation}
      \begin{equation}
	H^{r \theta_{1} \theta_{2}} = \mathcal{O}(r^{-4 - \epsilon})
\end{equation}
      \begin{equation}
	H^{\theta_{1} \theta_{2} \theta_{3}} = \mathcal{O}(r^{-5 - \epsilon})
\end{equation}
The condition that the magnetic charge (\ref{mcharge}) be finite 
actually imposes the stronger condition
\be
H^{\theta_{1} \theta_{2} \theta_{3}} = \mathcal{O}(r^{-6})
\ee
Any components of $ B$ of higher order than 
necessary to produce $ H$ are pure gauge and we choose a gauge 
where they do not appear.  The above fall-off are sufficient to show 
that  all 
the $\delta \pi$ and $\delta B$ terms  vanish.   Using again the 
finite 
energy requirement and the 
equation of motion for the dilaton we find:
\begin{equation}
    \phi =   C + \frac{a(\theta_{i})}{r^{1 + 
    \epsilon}} + \frac{b(\theta_{i},t)}{r^{3 + 
    \epsilon}}
    \end{equation}
    where $C$ is a constant. However, to obtain a finite asymptotic 
scalar charge one needs the stronger fall-off $\phi= C + 
\mathcal{O}(r^{-2})$.  If the perturbation is allowed to change the 
value of the constant at infinity, we get a scalar charge term, 
otherwise we  do not; these conclusions 
match those found by Gibbons, Kallosh, and Kol 
\cite{Gibbons:1996af}.  We will assume that the dilaton vanishes at 
infinity and hence there is no contribution from the matter fields to 
the first law. In particular, dipole charges do not seem to appear.

\setcounter{equation}{0}
\section{Emparan's dipole ring solutions}

We now briefly review the stationary black ring solutions to 
(\ref{theory})  found by 
Emparan \cite{Emparan:2004wy}.  (We follow Emparan's convention  
 and take $\beta = \frac{1}{16 \pi G}$ in the next two sections.)  The solutions 
depend on three parameters, but since only one 
component of the angular momentum is nonzero, there are only two 
conserved quantities, $M,J$. The third parameter is the dipole 
charge.  It is easiest to start with four auxiliary parameters 
$R,\lambda,\mu,\nu$ and later impose one constraint. These solutions 
are most conveniently expressed in terms of the following three 
functions 
\begin{equation}
	    F(\xi) = 1 + \lambda \xi, \quad  G(\xi) = 
	    (1-\xi^{2})(1+\nu \xi), \quad H(\xi) = 1- \mu \xi
	    \end{equation}
The black rings are independent of time, $t$, and two orthogonal 
rotations parameterized by $\varphi$ and $\psi$.  Introducing two 
other spatial coordinates, $-1\le x\le 1$ and $y\le -1$, the metric 
is:
$$ds^{2} = 
-\frac{F(y)}{F(x)}\Bigg(\frac{H(x)}{H(y)}\Bigg)^{\frac{N}{3}}\(dt + 
C(\nu,\lambda)DR \frac{1+y}{F(y)} d\psi\)^2
$$
\begin{equation}
   + \frac{R^{2}}{(x-y)^{2}}F(x) (H(x)H^{2}(y)\Big)^{\frac{N}{3}} 
    \Bigg [-\frac{D^{2}G(y)}{F(y)H^{N}(y)} d\psi^{2} 
    -\frac{dy^2}{G(y)} +\frac{dx^2}{G(x)} 
    +\frac{D^{2}G(x)}{F(x)H^{N}(x)} d\varphi^{2} \Bigg]
    \end{equation}
    while the dilaton is given by
    \begin{equation}
    e^{\phi} = \(\frac{H(x)}{H(y)}\)^{\frac{N\alpha}{2}}
    \end{equation}
    and the only nonzero component of the two form potential is 
    \begin{equation}
	B_{t\psi} = \frac{C(\nu, -\mu)\sqrt{N} DR (1+y)}{H(y)} + k
	\end{equation}
 $C$ and $D$ are given by	    $C(\nu, \lambda) = 
\sqrt{\lambda(\lambda - 
	    \nu)\frac{1+\lambda}{1-\lambda}}$ and $D 
=\frac{\sqrt{1 - \lambda} (1+\mu)^{\frac{N}{2}}}{1- \nu}$.   The 
horizon is at $y = -1/\nu$ with topology $S^{1}\times S^{2}$ 
where $\psi$ parametrizes the $S^{1}$ and $x$ and $\varphi$ 
parametrize 
the $S^{2}$.  The reader 
familiar with 
\cite{Emparan:2004wy}
 should note we take $\psi$ and $\vp$ to have period $2 
\pi$.  To avoid conical singularities along the $\varphi$-axis 
$x=\pm1$, one requires a relation between $\lambda,\mu,\nu$. 

 The $x,y$ coordinates 
break down near the axis and near infinity but making the following 
coordinate transformation one finds a manifestly asymptotically flat 
metric:
\begin{equation}
y = -1 -\frac{A \sin^{2}\theta}{r^2 + f(\theta)} \quad \quad x = 
-1 + \frac{A \cos^{2}\theta}{r^2 + f(\theta)} 
\end{equation}
where
\begin{equation}
A = \frac{2R^{2}(1 - \lambda)(1+\mu)^{N}}{1-\nu}
\end{equation}
and
\begin{equation}
f(\theta) = \frac{(1 - 3 \nu)A \cos^{2}\theta}{2(1 - \nu)} + c_{0}
\end{equation}
with $c_{0}$ an arbitrarily chosen constant.
Then asymptotically the dilaton is
\begin{equation}
    \phi =  -\sqrt{\Big(N - 
    \frac{N^2}{3}\Big)}\frac{\mu A}{(1+\mu)r^{2}} + 
    \mathcal{O}(\frac{1}{r^{4}})
    \end{equation}
while the potential asymptotically is:
\begin{equation}
B_{t \psi} = C(\nu, -\mu)D\sqrt{N}R\frac{1 + y}{1 - \mu y} + k = 
-C(\nu, -\mu)D\sqrt{N}R\frac{A\sin^{2}{\theta}}{(1 + \mu) r^{2}} + k+
\mathcal{O}(\frac{1}{r^{4}})
\end{equation}
where $C(\nu, -\mu) = \sqrt{\mu(\mu + \nu)\frac{1-\mu}{1+\mu}}$. 

It is easy to check that the dipole charge (\ref{localcharge}) is 
nonzero for this solution. The only angular momentum is in the $\psi$ 
direction.
Emparan computed the mass $M$, surface gravity $\kappa$, horizon area 
$A_{H}$, angular velocity $\Omega$, and angular momentum $J$ for 
these 
solutions and verified that they satisfy
   \begin{equation}\label{Emparan1stlaw}
	\delta M = \frac{\kappa}{8\pi G} \delta{A_{H}}  + \Omega \delta{J} + 
\phi_e
	\delta q_e
	\end{equation}
	where 
	\be\label{defphid}
	\phi_e= \frac{\pi}{2G} ( B_{t \psi}|_\infty - B_{t 
	\psi}|_{horizon})
	\ee  
In (\ref{Emparan1stlaw}), the perturbations are restricted to go from 
one stationary solution to another. But this is certainly included in 
the Sudarsky and Wald argument which applies to an arbitrary 
perturbation that satisfies the constraints. Since the dipole charge 
clearly appears in Emparan's first law,  we have an apparent  
contradiction. This is particularly puzzling since the dipole charge 
requires an integral over an $S^2$, and the Sudarsky-Wald derivation 
only produces integrals over the horizon and infinity which are 
three-surfaces.

 Let us turn now to the Hamiltonian formalism and explicitly evaluate 
the surface terms at infinity.  We take a
surface of constant $t$, a 
vector $n^{\mu}$ normal to these surfaces (and hence having nonzero 
$t$ 
and $\psi$ components), and $\xi^{\mu} = \chi^{\mu} = 
(\frac{\partial}{\partial t})^{\mu} + \Omega(\frac{\partial}{\partial 
\psi})^{\mu}$.  For Emparan's solutions, since the 
dilaton is independent of $ t$ and $\psi$, the momentum canonically 
conjugate to the 
dilaton vanishes.  The dilaton also goes to zero at infinity and so 
we 
get no scalar charge terms.  The non-vanishing components of the 
momentum 
conjugate to the 2-form 
${B}$ are
\begin{equation}
    \pi_{B}^{\psi r} = \frac{\sqrt{{h}}}{32 
\pi G} e^{-\alpha \phi} {n}_{\mu}
{H}^{\mu\psi r} = \frac{C(\nu, -\mu) 
D\sqrt{N}RA\sin\theta\cos\theta}{16\pi G (1 + \mu) r^2} + 
\mathcal{O}(\frac{1}{r^{4}})
\end{equation}
and
\begin{equation}
    \pi_{B}^{\psi \theta} = \frac{\sqrt{{h}}}{32 
\pi G} e^{-\alpha \phi} {n}_{\mu}
{H}^{\mu\psi \theta} = -\frac{C(\nu, 
-\mu)D\sqrt{N}RA\cos^{2}\theta}{16\pi G (1 + \mu) 
r^3} + \mathcal{O}(\frac{1}{r^{5}})
\end{equation}
These fields fall off sufficiently quickly to eliminate any surface 
terms at infinity.

\setcounter{equation}{0}
\section{Resolution}

The resolution to this apparent contradiction is an implicit 
assumption in the Sudarsky-Wald argument\footnote{To be fair, this 
condition was stated explicitly in \cite{Sudarsky:1992ty}, but its 
significance becomes 
clearer in the context of dipole charge.}:  the potential 
$B_{\mu\nu}$ must be globally defined and nonsingular everywhere 
outside (and on) the horizon. Since we are dealing with an electric 
dipole charge which does not have any obvious topological 
obstruction, this seems reasonable. However we now show that
this it is incompatible with our other assumptions that the dipole 
charge is nonzero and that $ B$ is invariant under the 
spacetime symmetries $\frac{\partial}{\partial t}$ and 
$\frac{\partial}{\partial \psi}$.  We first consider the case where 
the only 
angular velocity is  $\Omega_\psi$ (as in Emparan's solutions), and 
then comment on the generalization to $\Omega_\varphi \ne 0$.   First 
note that
$B_{\mu \psi}$ must vanish along the $\psi$-axis.  This is simply 
because $B_{\mu\psi} = B_{\mu\nu} (\partial/\partial \psi)^\nu$ and, 
by definition, $\partial/\partial \psi =0$ on the axis. If  
$B_{\mu\psi} \ne 0$, then $B_{\mu\nu}$ diverges, and this is not just 
a gauge effect. Set ${A_\nu} = \chi^\mu {B_{\mu\nu}} $.
Then a nonzero $\oint {A}\cdot dl$ for arbitrarily small loops 
around the $\psi$ axis indicates a $\delta $-function flux of $ 
H_{\mu\nu\rho} \chi^\rho$ along the axis (see eq. (4.3) below). This 
means that the constant 
$k$ in Emparan's solution for $B_{t \psi}$ is not arbitrary. In his 
solution, the $\psi$ axis is $y=-1$, and $k$ must be chosen so that 
$B_{t \psi}(y=-1)=0$.
However $B_{t \psi}$ must also vanish at the horizon 
\cite{Horowitz:1993wt}. This is because 
\be
B_{t \psi} = B_{\mu\nu} \chi^\mu \(\frac{\partial}{\partial 
\psi}\)^\nu
\ee
and $\chi^\mu=0$ on $S$. It is clearly impossible to satisfy both of 
these conditions in Emparan's solution. As presented in the previous 
section (and in \cite{Emparan:2004wy}) once $k$ is chosen to avoid a 
$\delta $-function flux of $ H$ along the $\psi$-axis, 
$B_{\mu\nu}$ necessarily diverges at the horizon.
Unlike the axis, this IS purely a gauge effect: the physical field 
$ H$ remains finite at the horizon. 

The inability to have $B_{t \psi}$ vanish at both the axis and 
horizon is
 not just a feature of Emparan's solution, but will be present 
whenever the dipole charge is nonzero.  Let us introduce a coordinate 
$y$ (as in Emparan's solution) so that constant $t,\psi,y$ label 
two-spheres which are continuously connected to the $S^2$ on the 
horizon. Aside from some factors involving the dilaton and metric, 
the dipole charge (\ref{localcharge}) involves an integral of $ 
H_{t\psi y}=\partial_y 
B_{t\psi} $ over $S^2$. If  $B_{t \psi}$ vanished at both the axis 
and horizon, then along every path connecting these surfaces, $ 
H_{t\psi y}$ would have to change sign. But  the charge is conserved 
and the factors of the dilaton and the metric in the integrand cannot 
change sign. Hence $B_{t \psi}$ cannot vanish on both the 
axis and horizon.

We will keep $ B_{\mu\nu}$ regular on the axis, but allow it to 
diverge 
on the horizon\footnote{We emphasize again that this is purely a 
gauge effect. One could use more than one patch and keep the 
potential finite everywhere, but the argument appears to be more 
complicated in this case.}. Consider again the boundary terms 
(\ref{bdyterms}) on $S$. We require the perturbations of the 
canonical variables to be finite on the horizon, so the vanishing of 
$N$ and $N^a$
on $S$ still cause all terms to vanish except
\be
2 \int_S dS_b  \chi^\mu  B_{\mu c}\delta\Bigg( \frac{ 
\pi_{B}^{bc}}{\sqrt{h}}\Bigg) 
\ee
Since $ B$ diverges, its contraction with $\chi$ can now 
remain nonzero on $S$. To evaluate this term, we use the fact that
\begin{equation} \label{lieidentity}
    d(\xi \cdot B) = \mathcal{L}_{\xi}B - \xi \cdot H
    \end{equation}
for any vector $\xi$ where a dot denotes contraction on the first 
index. 
If we take  $\xi = \chi$ then the right hand side vanishes on $S$ 
since $ B$ is invariant under $\chi$ and $\chi=0$ on $S$. So 
on the horizon  $\chi \cdot B$ 
 is a closed one form.  Hence it must be the sum of an exact form 
and a harmonic form. Since the $S^{2}$ in the horizon is simply 
connected, the only harmonic one form comes from the $S^1$. Hence 
\begin{equation}
    \chi\cdot B = df + cd\psi
    \end{equation}
    where $c$ is a constant.   The first term gives no contribution 
since  integrating the 
    surface term by parts and using the constraint 
(\ref{Bconstraint}) and the symmetry 
    of the extrinsic curvature (now using $\hat{n}^{a}$, the unit 
    normal to the horizon and to $n^{\mu}$) we see that it vanishes. 
     Using the fact that $ B$ is independent of $\psi$ we have 
$c = B_{t\psi} \vert_{horizon}$ and the surface term becomes
         \begin{equation}
	2c\int_S dS_{b}\delta \Bigg( \frac{ \pi_{B}^{b \psi}}{\sqrt{h}}\Bigg) = 
	\frac{c}{8G}\delta \int_{S^2} e^{-\alpha \phi} \star H = \frac{c 
\pi}{2G}  \delta q_e
	\end{equation}
Thus, the dipole charge does appear in the first law. Including the 
gravitational surface terms, we obtain 
        \begin{equation}\label{our1stlaw}
	\delta M = {\kappa\over 8\pi G} \delta{A}_{H} + \Omega_\psi 
\delta{J^\psi} + 
\phi_e \delta 
	q_e 
	\end{equation}
where $\phi_e =-\pi c/2G = -\frac{\pi}{2G} B_{t\psi}\vert_{horizon}$. This is 
identical to 
the first law found by Emparan, except for an apparent discrepancy in 
the definition of $\phi_e$.
However note that $B_{t\psi}$ must be constant over the sphere at 
infinity.  If not, $H_{t\psi\theta}$ would be nonzero asymptotically 
contradicting the falloff (\ref{falloff}). Since the $\psi$-axis goes 
off to infinity, the fact that $B_{t\psi}=0$ on this
axis implies that it must vanish everywhere at infinity. Thus our 
definition of $\phi_e$ 
indeed agrees with Emparan's  (\ref{defphid}). Note that the only 
symmetry of $ B$ that we needed to evaluate the surface term 
was that ${\cal L}_\chi  B=0$. To show that $B_{t\psi}$ could 
not vanish at both the horizon and the axis when the  dipole charge 
is nonzero, we used that $ B$ was independent of both $t$ and 
$\psi$ separately. We never needed to assume that $ B$ was 
independent of $\varphi$.

To summarize: the contradiction is resolved in two steps. The first 
is that the $B$ field cannot be finite at both the horizon and axis 
if the dipole charge is nonzero. Allowing $B$ to diverge on the 
horizon produces a nonzero surface term. The second step is to use 
the cohomology of the horizon together with its symmetry to show that 
the surface term is indeed related to the dipole charge. 

If $\Omega_\varphi\ne 0$, there are two changes to the first law 
(\ref{our1stlaw}). 
The obvious one is that one picks up a term $\Omega_\varphi 
\delta{J^\varphi} $ on the right hand side. The more subtle change is 
in the definition of the potential $\phi_e$. Since $\chi^\mu$ is now 
given by (\ref{defchi}), 
\be
\phi_e = -{\pi\over 2G} ( B_{t\psi} + \Omega_\varphi 
B_{\varphi \psi}) \Big\vert_{horizon}¥
\ee 
Solutions have not yet been found in which black rings with dipole 
charge have nonzero angular velocity in both the $\varphi$ and $\psi$ 
directions. However, there is no reason why they should not exist.

 \setcounter{equation} {0}
  \section{Local Charges and Minimal 5D Supergravity}

 Recently, a family of black ring solutions were found in minimal 5D 
supergravity \cite{Elvang:2004xi}. Unlike Emparan's original dipole 
rings discussed in section 3, these solutions were sufficiently 
complicated that a first law could not be found by 
inspection.\footnote{Larsen \cite{Larsen:2005qr} has recently given a 
first law based on a model for the microscopic degrees of freedom. 
However this only applies to the near extremal solutions and is 
formulated in terms of near horizon quantities (mass, charges, etc.) 
which do not agree with the usual asymptotically defined 
quantities.}  The general procedure of Sudarsky and Wald can be used 
to derive a first law. Our earlier analysis does not immediately 
apply to this case since  the supergravity action contains a 
Chern-Simons term which we have not included. In this 
section we extend our analysis to include this term.   Specifically, 
we consider the five dimensional action
\begin{equation}
    S = \beta \int d^5 x\sqrt{-g} \Big(R -\frac{1}{4} F_{\mu \nu} 
F^{\mu \nu} - \gamma 
    \epsilon^{\mu \nu \rho \sigma \eta} F_{\mu \nu} F_{\rho \sigma} 
A_{\eta} \Big)
    \end{equation}
   Minimal 5D supergravity corresponds to $\gamma = (12\sqrt 
3)^{-1}$. (If we set $\gamma=0$, this theory is equivalent to 
(\ref{theory}) with $\alpha=0$ and the dilaton removed.)  Before 
discussing the 
    details it is worth pointing out an important difference between 
this case and the discussion in 
     section 2.  Since we are now working with a two form $F$ rather 
than a three form $H$, the dipole charge is a magnetic charge
    \be
    q_m= {1\over 4\pi} \int_{S^2} F
    \ee
     A nonzero dipole charge clearly implies that we will not have a 
potential which 
    is globally defined, so we must work in patches. However, in each 
patch, we can choose $A$ so that it is finite on the horizon.  
      Hence when we evaluate the surface terms from the variation of 
the Hamiltonian, we will have no Maxwell contributions from 
    the horizon of the black hole. The dipole charge will  appear 
through surface terms on the interface between the patches.
    
    The Chern-Simons term clearly enters the field equation for $F$ 
which now takes the form
    \be\label{FeqCS}
    \nabla _\mu F^{\mu\eta}  -3\gamma\epsilon^{\mu \nu \rho \sigma 
\eta} F_{\mu \nu}
    F_{\rho\sigma} =0
    \ee
Since the Gauss law constraint is just the time component of this 
equation, it too will be modified by the Chern-Simons term. 
However  this term is independent of the spacetime metric, and so 
does not contribute to the stress energy tensor. Since the 
gravitational constraints involve the matter only through components 
of $T_{\mu\nu}$, one might expect that the gravitational constraints 
are unaffected by the Chern-Simons term. Unfortunately, this is not 
the case. The Hamiltonian must be expressed in terms of the canonical 
momenta and the momentum 
    conjugate to $A$ now has a Chern-Simons contribution:
    \begin{equation}\label{csmom}
	\pi^{a} = \beta \sqrt{h} (F^{\mu a}n_{\mu} + 4 \gamma 
\epsilon^{abcd} F_{bc} A_{d} )
	\end{equation}
where we define the four dimensional $\epsilon^{abcd} = 
\epsilon^{abcd\mu}n_{\mu}$.
	Note, in particular, that since ${A}$ is only defined in patches, so 
	is $\pi^{a}$.  After computing the canonical Hamiltonian, we find that 
it has a pure 
	constraint form, as we expect on general grounds:
	\begin{equation}
	    \mathcal{H}_{V} = \xi^{\mu} C_{\mu}¥ + 
	    \xi^{\mu} A_{\mu} \mathcal{C}
	    \end{equation}
	    where  the general relativity constraints $C_{\mu}$ and 
	     the gauge constraint $\mathcal{C}$ both contain Chern-Simons 
contributions.  Explicitly
  \begin{equation}
		\mathcal{C} = -\beta \sqrt{h}\Bigg(D_{a}\Big( \frac{\pi^{a}}{\beta 
\sqrt{h}} \Big) + 
		\gamma \epsilon^{abcd} F_{ab} F_{cd} \Bigg)
		\end{equation}
 $$
C_{0} = -2 \sqrt{h} (G_{\mu \nu} - 8\pi T_{\mu \nu})n^{\mu}n^{\nu}   
= - \beta \sqrt{h} R^{(4)} 
+ \frac{1}{\beta \sqrt{h}}(\pi_{G}^{ab}\pi^{G}_{ab} - 
\frac{\pi_{G}^2}{3})
$$
\begin{equation}
+ \frac{\pi^{a} \pi_{a}}{2 \beta \sqrt{h}} + \frac{\beta \sqrt{h}}{4} 
F_{ab} F^{ab} - 4\gamma \epsilon^{abcd} \pi_{a} {F}_{bc} A_{d} + 8 
\gamma^{2} \beta \sqrt{h} 
\epsilon^{abcd}\epsilon_{ajkl} F_{bc} A_{d} F^{jk} A^{l} 
\end{equation}
\begin{equation}
C_{a} =  -2\sqrt{h} (G_{a\mu} - 8\pi T_{a\mu})n^{\mu}  
=  -2\sqrt{h}{h}_{{ab}}D_{{c}} ( 
\frac{\pi_{G}^{bc}}{\sqrt{h}}  ) + 
F_{ab} (\pi^{b} - 4\gamma \beta \sqrt{h} \epsilon^{bcde} F_{cd} A_{e})
\end{equation}
Much of the complication in these expressions arises from replacing 
the electric field $E^a = F^{\mu a} n_\mu$ which appears in 
$T_{\mu\nu}$ by the canonical momentum (\ref{csmom}). This is 
necessary since to derive the appropriate surface terms we must vary 
the Hamiltonian with respect to the canonical variables. Requiring 
that the Hamiltonian have a well defined variation leads to the 
following surface terms (in addition to the usual gravitational  
terms)  
$$    \beta \int dS_{b} \Bigg[ \xi^{\mu} A_{\mu} \delta 
    \Big(\frac{\pi^{b}}{\beta \sqrt{h}} \Big) + \Big(N F^{ab} - 2 
E^{[a}N^{b]}\Big)\delta A_{a} 
   $$
    \begin{equation} \label{CSsurface}
	+4\gamma \epsilon^{bcda}\Big(\xi^\mu A_\mu F_{cd} - 2 
F_{c\mu}\xi^{\mu} A_{d} \Big) \delta A_{a} \Bigg]
	\end{equation}
 We now discuss the interpretation of these surface terms.

Suppose we have a black ring solution which is stationary and 
axisymmetric (in both orthogonal planes) and has a bifurcate Killing 
horizon. As before, let $\Sigma$ be a spacelike surface which is 
asymptotically flat and has an inner boundary at the bifurcation 
surface $S$. Also, set the time evolution vector, $\xi^\mu$, equal to 
the Killing field $\chi^\mu$ which is tangent to the horizon and 
vanishes on $S$. There are no contributions from the inner boundary 
since $\xi^\mu=0$ there. At infinity, our finite energy conditions 
ensure that only the first term contributes and we get the usual 
$\Phi_{E} \delta Q_{E}$ global electric charge term. To see this, 
recall that in this theory, the definition of a conserved electric 
charge depends on the Chern-Simons term in general. From the 
constraint, its clear that
\be
Q_E = {1\over 4\pi} \int_{S^3} dS_a \Big[ \frac{\pi^{a}}{\beta 
\sqrt{h}} + 2\gamma \epsilon^{abcd} A_b F_{cd} \Big]
\ee
is independent of which $S^3$ it is evaluated on. However, with 
standard asymptotically flat boundary conditions, the Chern-Simons 
term does not contribute to the charge computed at infinity. Also, in 
five dimensions, $A_\vp$ and $A_\psi$ must vanish asymptotically, so 
$\Phi_E = -  4\pi \beta \chi^\mu A_\mu \Big \vert_{r = \infty} = -4\pi \beta A_t\Big 
\vert_{r = \infty}$.

The new complication arises from the fact that we have  
magnetic dipole charge associated with the $S^{2}$ of the horizon.  
This means that we must divide our surface $\Sigma$ into two patches. 
The surface terms (\ref{CSsurface}) will then arise on the interface 
between the two patches.  Each patch produces surface terms of the 
same form (with their appropriate $A$), but with opposite sign, so we 
are interested in the difference between these two contributions. The 
new  black ring solutions \cite{Elvang:2004xi} were found in C-metric 
like coordinates similar to those used in Emparan's dipole rings. So 
we 
again use coordinates like those in the previous section 
$(x,y,\vp,\psi)$, where $-1\le x\le 1$ and $\vp$ parameterize an 
$S^2$, $\psi$ parameterizes the 
 $S^{1}$ of the horizon, and $y$ is like a radial coordinate. We 
choose $-1 < x_0 <1$ and define our two patches to be $x<x_0$ and 
$x>x_0$.  $A_\vp$ is discontinuous across $x=x_0$ with $\Delta A_\vp 
= 2q_m$. The surface $x=x_0$  begins and ends on the horizon. It does 
not enter the asymptotic region.

We have previously assumed that 
 $\xi^{\mu} 
A_{\mu}$ vanishes at the horizon, so  we have no contributions there.
  To do this, however, in the presence of a magnetic dipole 
charge we need to be able to set $\xi^{\mu} A_{\mu}$ to zero in both 
patches. In particular,  $\xi^{\mu} A_{\mu}$ must be continuous 
across 
the interface between patches. If $\Omega_\vp\ne 0$  this means that 
$A_{t}$ is discontinuous by 
$2 \Omega_{\vp} q_{m}$.  (We will work in a gauge in which $A_\psi$ 
is continuous.) For all presently known nonextremal black rings with 
dipole charges, 
$\Omega_\vp=0$,  
but this does not seem to be a fundamental restriction and we expect 
solutions with non-vanishing $\Omega_\vp$ to be discovered in the 
future. Since the interface between our two patches does not enter 
the asymptotic region,  $A_{t}$ is continuous at large radius and 
there is no ambiguity in the $\Phi_E
\delta Q_E$ term. 

We can now evaluate the contribution from the surface terms on the 
interface between our two patches.  At first sight, it appears that 
there will be terms proportional to the dipole charge $q_{m}$ and 
not just its variation. Fortunately, those terms cancel. After a 
bit of algebra we find all the terms in 
(\ref{CSsurface}) reduce down to a surprisingly simple $\phi_{m} 
\delta q_m$ where we define
\begin{equation}\label{CSpot}
    \phi_{m} = - 2\beta \int dS_{b} \Big[ N F^{b c} - 2
    E^{[b} N^{c]} - 12 \gamma  \xi^{\mu} 
    A_{\mu} \epsilon^{bcde}
     F_{de} \Big] D_c\vp
    \end{equation}
where the integral is over the interface between the patches. To make 
this more covariant, one could replace the $D_c \vp$ with 
$\frac{\Delta A_c}{2 q_{m}}$. To be well defined, this potential must 
not change when we deform the surface of integration, since this just 
corresponds to choosing different gauge patches. In other words, the 
divergence of the integrand must vanish. This is far from obvious, 
but we have checked that the potential is indeed independent of 
surface whenever $A_\mu$ satisfies the field equation (\ref{FeqCS}) 
and 
${\cal L}_\xi A =0$.  This provides a highly nontrivial check of this 
potential.

The net result is a standard looking first law 
 \be
\delta M = \frac{\kappa}{8\pi} \delta{A}_{H} + \Omega_{i} 
\delta{J}^{i} + \Phi_E \delta Q_E+ \phi_m 
\delta q_m 
 \ee
 where the magnetic dipole potential is given by (\ref{CSpot}).	

\setcounter{equation}{0}
 \section{Generalization to higher dimensions}

Having seen that dipole charges can appear in the first law, we now 
investigate whether other charges might arise. For simplicity, we 
will drop the Chern-Simons term and consider a higher dimensional  
generalization of (\ref{theory}) including  a $p$-form potential and 
dilaton in $d$ dimensions
 \begin{equation}
     S = \beta \int d^dx \sqrt{-g} \left [ {R} 
-\frac{1}{2}
\nabla_\mu
     \phi \nabla^\mu\phi - \frac{1}{2({p} + 1)!}e^{-\alpha
     \phi} H_{\mu_1 \cdots \mu_{p+1}}H^{\mu_1 \cdots \mu_{p+1}} 
\right ]
     \end{equation}
     where $\phi$ is the dilaton and ${H} = {dB} $ is a
     $(p + 1)$-form field strength.  Then let us perform the usual 
Hamiltonian decomposition with $N$ the lapse function, $N^{a}$ 
the shift vector, ${h}_{ab}$ the induced metric on a
surface $\Sigma$ of constant time and $n^{\mu}$ the unit 
normal to  $\Sigma$.
 The momentum canonically conjugate to the spatial metric ${h}_{ab}$ 
and dilaton $\phi$ are again given by (\ref{piG}) and (\ref{piphi}) 
respectively. The momentum conjugate to the $p$-form potential 
$ {B}$ is
\begin{equation}
\pi_{B}^{a_{1} \ldots a_{p}} = 
\frac{\partial
\mathcal{L}}{\partial
\dot{B}_{a_{1} \ldots a_{p}}} = \frac{\beta \sqrt{h}}{p!} e^{-\alpha 
\phi} {n}_{\mu}
{H}^{\mu a_1 \ldots a_p}
\end{equation}

\indent  We define the Hamiltonian volume density 
\begin{equation}
\mathcal{H}_{V} = \pi_{G}^{ab}\dot{{h}}_{ab} + 
\pi_{B}^{a_{1} \ldots a_{p}}\dot{B}_{a_{1} 
\ldots a_{p}} + \pi_{\phi}\dot{\phi}  - \mathcal{L}
\end{equation}
where $\mathcal{L}$ is the Lagrangian density.  Then performing 
integrations by parts to put the result in pure 
constraint 
form we obtain
\begin{equation}
{H}_{V}  
 = \int_\Sigma \Big(\xi^{\mu}C_{\mu} + \xi^{\mu_{1}}B_{\mu_{1} 
\mu_{2} \ldots 
\mu_{p}}\mathcal{C}^{\mu_{2} \ldots \mu_{p}}\Big)
\end{equation}
where $C_{\mu}$ is the constraint from the Einstein equations and $ 
\mathcal{C}^{\mu_{2} \ldots \mu_{p}} $ is the constraint from the 
$p$-form.  Explicitly,
$$
C_{0} = -2\sqrt{h} (G_{\mu\nu} - 8\pi T_{\mu\nu})n^{\mu}n^{\nu}   
= - \beta \sqrt{h} R^{({d} - 1)} 
+ \frac{1}{ \beta \sqrt{h}}(\pi_{G}^{ab}\pi^{G}_{ab} + 
\frac{\pi_{G}^2}{2 - {d}})
$$
\begin{equation}
+ \frac{\pi_{\phi}^2}{2 \beta \sqrt{h}} + \frac{\beta \sqrt{h}}{2}(D 
\phi)^2 + \frac{p!}{2 \beta \sqrt{h}} e^{\alpha \phi} \pi_{B}^2 + 
\frac{\beta \sqrt{h}}{2 (p+1)!} e^{-\alpha \phi} H_{a_{1} \ldots a_{p+1}} 
H^{a_{1} \ldots a_{p+1}}
\end{equation}
\begin{equation}
C_{a} =  -2\sqrt{h} (G_{a \mu} - 8\pi T_{a \mu}) n^{\mu}  
=  -2\sqrt{h}D_{c} ( 
\frac{{\pi_{G}}^{c}_{a}}{\sqrt{h}}  ) + 
\pi_{\phi}D_{a}\phi 
+  \pi_{B}^{a_1 \ldots a_{p}} H_{a a_{1} \ldots a_{p}}
\end{equation}
\begin{equation}
\mathcal{C}^{a_{2} \ldots a_{p}} = -p\sqrt{h} D_{a}\Big(\frac{\pi_{B}^{a 
a_{2} \ldots a_{p}}}{\sqrt{h}}\Big)
\end{equation}

\indent Now the variation of ${H_{V}}$ is well defined only if 
one adds appropriate surface terms. In addition to the usual 
gravitational terms, we must add
$$
- \beta \int dS_{a_1} \Big[ \Big( N D^{a_1} \phi + 
N^{a_1} \frac{\pi_{\phi}}{\beta \sqrt{h}} \Big) \delta \phi - p 
\xi^{\mu} B_{\mu 
a_{2} \ldots 
a_{p}}\delta\left( \frac{ \pi_{B}^{a_1 \ldots a_{p}}}{\beta 
\sqrt{h}}\right)
$$
\begin{equation}\label{delta H}
+ \Big( \frac{N  e^{-\alpha \phi}}{p!} H^{a_{1} 
\ldots a_{p + 1}}+ \frac{p + 1}{\beta \sqrt{h}} N^{[a_{1}} 
\pi_{B}^{a2 \ldots a_{p + 1}]}\Big) \delta B_{a_{2} \ldots 
a_{p+1}} \Big]
\end{equation}

\indent Let us again specify what we mean by asymptotically flat.  We 
take 
the metric to be flat with order $\frac{1}{r^{d - 3}}$ corrections, 
as in the Myers-Perry black holes.  
We again require the solutions to have finite energy and hence 
$T_{\mu\nu}n^{\mu}n^{\nu} = \mathcal{O}(r^{1 - d - \epsilon} )$.   At 
leading 
order $T_{\mu\nu}n^{\mu}n^{\nu}$ is given by 
    a sum of positive definite terms and hence we get the following 
    restrictions:
    \begin{equation}
	H^{t r \theta_{1} \ldots \theta_{p - 1}} = 
\mathcal{O}(r^{\frac{3-d}{2} - 
    p - \frac{\epsilon}{2}})
\end{equation}
     \begin{equation}
	H^{t \theta_{1} \ldots \theta_{p}} = \mathcal{O}(r^{\frac{1-d}{2} - 
    p - \frac{\epsilon}{2}})
\end{equation}
      \begin{equation}
	H^{r \theta_{1} \ldots \theta_{p}} = \mathcal{O}(r^{\frac{1-d}{2} - 
    p - \frac{\epsilon}{2}})
\end{equation}
      \begin{equation}
	H^{\theta_{1} \ldots \theta_{p+1}} = \mathcal{O}(r^{-\frac{d+1}{2} - 
    p - \frac{\epsilon}{2}})
\end{equation}
This is sufficient to ensure that all 
the $\delta \pi$ and $\delta B$ terms vanish except the last one. 
Previously, we used the fact that the magnetic charge (\ref{mcharge}) 
must be finite to argue that this term must also vanish. But this  
charge is only defined when $p=d-3$.  For now we simply 
 assume here that the last term also falls off sufficiently 
rapidly to give  no contribution at infinity.  Using again the finite 
energy requirement and the 
equation of motion for the dilaton we find:
\begin{equation}
    \phi \rightarrow C + \frac{a(\theta_{i})}{r^{\frac{d-3}{2} + 
    \frac{\epsilon}{2}}} + \frac{a(\theta_{i},t)}{r^{\frac{d+1}{2} + 
    \frac{\epsilon}{2}}}
    \end{equation}
    where $C$ is a constant.  To have a well defined scalar charge, 
we require the
    faster fall-off: $\phi = C + \mathcal{O}(r^{3-d})$. If the 
perturbation is allowed to change the constant value of $\phi$, there 
is a scalar charge term in the first law, otherwise, there is not:  
these conclusions 
match those found by Gibbons, Kallosh, and Kol. 
\cite{Gibbons:1996af}. We will assume that $\phi$ vanishes 
asymptotically, so the scalar surface terms vanish.

 Let us now consider generalizations of the first law  in section 
four.  The simplest generalization  occurs when $p=2$ (as in our 
previous example). Suppose there exists a black ring with horizon 
topology
 ${\mathcal{M}} \times 
S^{1}$ where ${\mathcal{M}}$ denotes any simply connected $d-3$ 
dimensional 
manifold.   Since  $\star{H}$ is a $d-3$ dimensional form, one can 
again define
 a dipole charge, $q_e \propto
\int_{{\mathcal{M}}} e^{-\alpha\phi} \star H$, and proceed as before. 
An essentially identical argument yields the first law:
        \begin{equation}
	\delta M = \frac{\kappa}{8\pi} \delta{A}_{H} + \Omega_{i} 
\delta{J}^{i} + \phi_e 
\delta 
	q_e
	\end{equation}
	where $\phi_e \propto B_{t\psi}$ evaluated on the horizon (and 
$\psi$ is the coordinate along the $S^1$). 
  
For $p>2$, the  general situation is the following. Suppose there is 
a bifurcate Killing horizon $S$ and let $\chi$ be the Killing field 
which vanishes on $S$.   The horizon can be an arbitrary $d-2$ 
dimensional manifold, but we will assume there is a nontrivial 
(torsion-free) 
$d-p-1$ cycle $T$. Then we can define a local charge (it is no longer 
a dipole charge) by
\be
q_l \propto \int_T  e^{-\alpha\phi} \star H
\ee
It seems likely that it will again be impossible to find a  potential 
$B$ which is finite
 and globally defined outside the horizon, and invariant under 
$\chi$. In this case, the surface term 
 \be\label{pdst}
 p\int_S \xi^{\mu} B_{\mu a_{2} \ldots 
a_{p}}\delta \left ( \frac{ \pi_{B}^{a_1 \ldots 
a_{p}}}{\sqrt{h}}\right ) dS_{a_1}
\ee
can be nonzero.  To evaluate this term, it is convenient to note that 
it is proportional to
\be
 \int_S  (\chi\cdot  B) \wedge \delta(e^{-\alpha\phi} 
\star  H)
\ee
where the dot denotes contraction on the first index of $B$.
 Applying (\ref{lieidentity}) we again see that $\chi\cdot  B$ 
is a closed $p-1$ form on the horizon, so it can be written as the 
sum of an exact and a harmonic form. An exact form does not 
contribute since $d(e^{-\alpha\phi} 
\star  H) =0$ by the field equation. (Actually, all we need is the 
constraint, which is the spatial projection of this equation.) So the 
only contribution comes from the harmonic part of $\chi\cdot  B$.
By the usual duality between homology and cohomology,  there is a 
harmonic form $\omega$ which is dual to $T$ in the sense that for any 
$d-p-1$ 
form $\sigma$
\be
\int_T  \sigma = \int_S \sigma \wedge \omega
\ee
It then follows that the surface term takes the form $\phi_l \delta 
q_l$ where the potential $\phi_l$ is just the constant relating the 
harmonic part of $\chi\cdot  B$ to $\omega$.
The first law will then include this new local charge.

\setcounter{equation}{0}	
\section{Discussion}

We have resolved an apparent contradiction between a general 
derivation of the first law for higher dimensional black holes and an 
explicit family of solutions found by Emparan.
The resolution is based on the realization that there does not exist 
a globally defined, nonsingular two-form potential $ B$ which 
respects   
the symmetry. This should not come as a surprise. Even in four 
dimensional Einstein-Maxwell theory one has a similar situation when 
considering more than one extremal black hole. Here again, one cannot 
find a globally defined potential $A_\mu$ which is static and 
finite on all of the horizons. For the five dimensional  rotating 
black ring, one has a problem even for a single object since the 
rotation axis plays the role of one of the horizons in the sense that 
it imposes a constraint on the behavior of the potential.

We have also derived a first law for black ring solutions in minimal 
5D supergravity. The Chern-Simons term produces an extra complication 
in the analysis, but in the end, a standard first law is obtained 
with a nontrivial potential for the magnetic dipole charge.
It is likely that there also exist asymptotically flat black holes 
with horizon topology other than $S^n$ or $S^1\times S^2$. We have 
discussed some local charges that these holes might carry and their 
contribution to the first law.

Our derivation of the first law requires a bifurcate Killing horizon. 
Hence it does not immediately apply to extremal black holes such as 
the 
 supersymmetric black rings 
\cite{Elvang:2004rt,Gauntlett:2004wh,Bena:2004de}.  However, the 
matter surface terms we derived are generic and could be used to derive a first law even in the absence of a bifurcation surface.

There are several possible generalizations of our work. We have so 
far considered just a single dipole charge. It should be 
straightforward to extend this to include several dipole charges and 
several global charges such as those which arise in dimensional 
reductions of ten and eleven dimensional  supergravity.
In four dimensions, the first law has been generalized to apply to 
isolated horizons in non-stationary spacetimes \cite{Ashtekar:2004cn}. 
This has also been discussed in the context of higher dimensions 
\cite{Korzynski:2004gr}, but without local charges.
It would be interesting to extend our derivation of the first law 
with dipole charges to 
this case.

\vskip 1cm
\centerline{\bf Acknowledgements}

\vskip .5cm
It is a pleasure to thank D. Marolf and R. Wald for discussions. This 
work was supported in part by
NSF grant PHY-0244764.

\end{document}